# Scaling and High-Frequency Performance of AlN/GaN HEMTs

Xi Luo[1], Subrata Halder[1], Walter R. Curtice[1], James C. M. Hwang[1], Kelson D. Chabak[2], Dennis E. Walker, Jr.[2], and Amir M. Dabiran[3]

[1]Lehigh University, Bethlehem, PA 18015, USA
[2]Air Force Research Laboratory, Wright Patterson AFB, OH 45433, USA
[3]SVT Associates, Inc., Eden Prairie, MN 55344, USA
Email: xil609@lehigh.edu

*Abstract* — **Small- and large-signal RF characteristics were measured on AlN/GaN HEMTs with 80-160 nm gate length and 100-300 μm width. Consistent with the literature, current-gain cut-off frequency and maximum frequency of oscillation were found to increase with inverse gate length and independent of gate width. For the first time, output power and efficiency were reported at the high end of X-band, and were comparable to the best reported at 2 GHz and insensitive to gate length or width. These results suggest that the AlN/GaN HEMTs can be further scaled for even higher frequency and higher power performance.**

*Index Terms* — **Cutoff frequency, gallium compounds, HEMTs, microwave transistors, pulse measurement, power measurement.**

## I. INTRODUCTION

Of all high-electron-mobility transistors (HEMTs), the AlN/GaN HEMT has the highest carrier density and adequate mobility, which result in the highest channel conductance [1]. With AlN as thin as a few nanometers, the AlN/GaN HEMT has also the highest mutual transconductance and can be aggressively scaled for millimeter-wave power applications without short-channel effects [2]. With record-setting DC characteristics of AlN/GaN HEMTs, about two-dozen papers have been published [3]-[26] since 2006 on small-signal characteristics such as the current-gain cut-off frequency $f_T$ and the maximum frequency of oscillation $f_{MAX}$ (Table I). Fig. 1 shows the scaling of $f_T$ and $f_{MAX}$ with the gate length $L_G$. In general, $f_T$ increases with inverse $L_G$ with $f_T \cdot L_G \approx$ 10±6 GHz·μm. However, scaling of $L_G$ was systematically investigated only in [3] and [7], and the gate width $W_G$ was always fixed. Large-signal characteristics were reported in [9], [10], [13], [17] only at an operating frequency $f$ = 2 GHz, with saturated output power $P_{SAT}$ = 0.5-2.6 W per mm gate width at a drain-source voltage $V_{DS} \approx$ 15 V. This paper further explores scaling of both $L_G$ and $W_G$, especially concerning their effects on small- and large-signal RF characteristics at X-band.

## II. EXPERIMENTAL

The present AlN/GaN heterostructure was grown on a c-plane sapphire substrate by plasma-assisted molecular beam epitaxy with a 0.7-μm-thick AlGaN buffer layer, a 2-μm-thick GaN channel layer, a 3.5-nm-thick AlN barrier layer, and a 1-nm-thick GaN cap layer. The combination of GaN and its native oxide helped passivate the surface and suppress gate leakage. (100-nm-thick SiN passivation was added beside the gate.) The gate metal NiAu was patterned by electron-beam lithography into gates of $L_G$ = 80 nm, 120 nm, and 160 nm. Each gate was centrally located in a 2.7-μm spacing between source and drain Ti/Al/Ni/Au metals. Each HEMT contained two gate fingers with $W_G$ = 50 × 2 μm, 100 × 2 μm, or 150 × 2 μm. Details of material growth and device processing can be found in [1] and [15], respectively.

TABLE I
RF CHARACTERISTICS OF ALN/GAN HEMTS

| $L_G$ (μm) | $W_G$ (μm) | $f_T$[a] (GHz) | $f_{MAX}$[a] (GHz) | $f$ (GHz) | $P_{SAT}$ (W/mm) | PAE (%) | Reference |
|---|---|---|---|---|---|---|---|
| 0.06-0.25 | 50×2 | 83-31 | 139-89 | | | | [3] |
| 0.1 | 50×2 | 87 | 149 | | | | [4] |
| 0.2 | 10 | 52 | 60 | | | | [5], [8] |
| 0.25 | 15 | 24 | 52 | | | | [6] |
| 0.05-0.2 | 50×2 | 106-49 | | | | | [7] |
| 1 | 200 | 6 | 11 | 2 | 0.85 | 24 | [9], [10] |
| 1 | 200 | 10 | 32 | | | | [11], [16] |
| 0.25 | 2 | 60 | 3.5 | | | | [12] |
| 1.3 | 150 | 9 | 32 | 2 | 2.6 | 33 | [13] |
| 0.4 | 200 | 20 | 31 | | | | [14] |
| 0.15 | 150×2 | 25 | 22 | | | | [15] |
| 0.15 | 50 | 50 | 102 | 2 | 0.53 | 41 | [17] |
| 3 | 100 | 3 | 8 | | | | [18] |
| 0.4 | 200 | 20 | 37 | | | | [19] |
| 0.08 | 38×2 | 112 | 215 | | | | [20] |
| 0.2 0.5 | 100 | 80 40 | 65 55 | | | | [21], [26] |
| 0.15 | 25×2 | 75 | 115 | | | | [22] |
| 0.15 | 38×2 | 82 50 | 210 150 | | | | [23] |
| 0.2 | 100 | 50 | 40 | | | | [24] |
| 0.16 | 50 | 85 | 103 | | | | [25] |
| 0.08 | 50×2 | 68 | 103 | | 2.7 | 31 | This Work |
| 0.12 | 50×2 | 57 | 82 | | 2.4 | 32 | This Work |
| 0.16 | 50×2 | 50 | 76 | | 2.2 | 29 | This Work |
| 0.08 | 100×2 | 72 | 114 | 12 | 2.6 | 30 | This Work |
| 0.12 | 100×2 | 63 | 96 | 12 | 2.3 | 33 | This Work |
| 0.16 | 100×2 | 51 | 77 | | 2.1 | 31 | This Work |
| 0.08 | 150×2 | 72 | 97 | | 2.3 | 26 | This Work |
| 0.12 | 150×2 | 55 | 90 | | 2.1 | 27 | This Work |
| 0.16 | 150×2 | 50 | 80 | | 2.0 | 24 | This Work |

[a]Extrinsic values from as-measured *S*-parameters whenever possible.

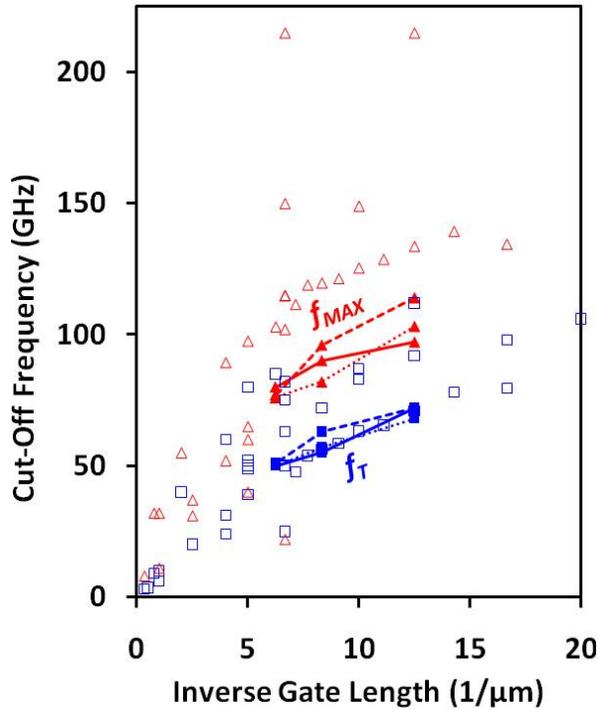

Fig. 1. Published (□) $f_T$ and (△) $f_{MAX}$ of AlN/GaN HEMTs vs. measured (■) $f_T$ and (▲) $f_{MAX}$ in this work. $W_G$ = (· · ·) 50 × 2 μm, (- - -) 100 × 2 μm, or (——) 150 × 2 μm.

The HEMTs were characterized on-wafer. Small-signal characteristics were measured from 6 to 40 GHz in a 50-Ω system. The resulted S-parameters were used to extract $f_T$, $f_{MAX}$, and other equivalent-circuit parameters [27]. Large-signal characteristics were measured at $f$ = 12 GHz under optimum input and output matches. To minimize heating, both $V_{DS}$ and the RF input power $P_{IN}$ were turned on for only 1 μs with a repetition frequency of 2.78 KHz. RF output power $P_{OUT}$ was sampled in the middle of the pulse.

### III. DC AND SMALL-SIGNAL CHARACTERISTICS

Van der Pauw measurement showed that the present AlN/GaN HEMTs had a carrier density of ~3 × 10$^{13}$ cm$^{-2}$ and a carrier mobility of ~800 cm$^2$/V·s, which resulted in a sheet resistivity of ~300 Ω/□. Transmission-line measurement confirmed that the sheet resistivity was 279±31 Ω/□. Typically, the HEMTs exhibited a pinch-off voltage of approximately −2 V with a subthreshold slope of ~0.5 V/decade, a saturated drain current of ~500 mA/mm at $V_{GS}$ = 0, and an extrinsic transconductance $G_M$ ≈ 500 mS/mm. The off-state drain-source breakdown voltage was approximately 20 V; the off-state drain-gate breakdown voltage was approximately 40 V. The gate leakage current was <100 μA/mm. The contact resistance was ~0.5 Ω·mm; the total source resistance was ~1 Ω·mm;

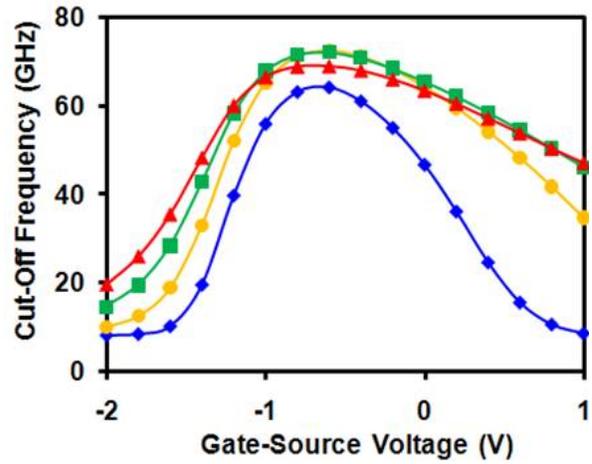

Fig. 2. Measured $f_T$ of present AlN/GaN HEMT. $L_G$ = 80 nm. $W_G$ = 100 × 2 μm. $V_{DS}$ = (♦) 2 V, (■) 4 V, (●) 6 V, or (▲) 8 V.

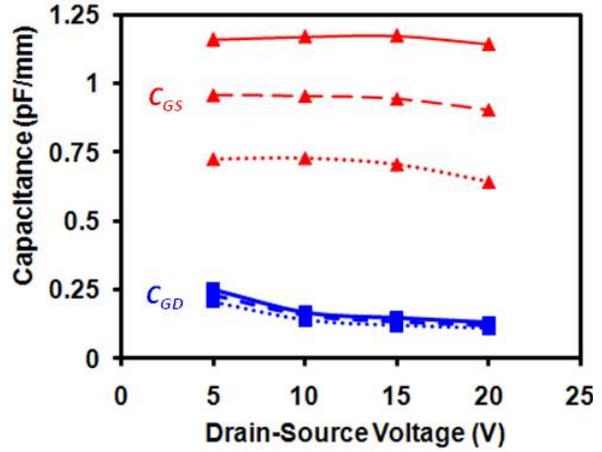

Fig. 3. Measured (▲) $C_{GS}$ and (■) $C_{GD}$ of present AlN/GaN HEMTs. $L_G$ = (· · ·) 80 nm, (- - -) 120 nm, or (——) 160 nm. $W_G$ = 150 × 2 μm. $V_{GS}$ = −0.2 V.

the total channel and drain resistance was ~2 Ω·mm; the gate resistance was ~5 Ω·mm. Detailed discussions of the DC characteristics can be found in [28].

Fig. 1 shows also that $f_T$ and $f_{MAX}$ of the present HEMTs increase with inverse gate length and follow closely the trend established by the pioneering work of [3], except $f_{MAX}$ is approximately 20% lower probably due to higher output conductance caused by higher buffer leakage [28]. There is no significant dependence on gate width, which suggests that up to 150 μm per finger width is tolerable up 40 GHz. Fig. 2 shows that typical of all present HEMTs, $f_T$ peaks around $V_{GS}$ = −0.6 V and is fairly constant for $V_{DS}$ > 2 V − the knee voltage $V_{KNEE}$. Fig. 3 shows that, although the gate capacitance decreases with decreasing gate length, there is a fixed parasitic component on the order of 0.1 pF/mm. This will eventually cause gate length scaling to reach diminishing return. As expected, the gate-source capacitance $C_{GS}$ is relatively independent of $V_{DS}$, while the gate-drain capacitance $C_{GD}$ decreases rapidly with increasing $V_{DS}$. Fig. 4 shows that the extracted AC transconductance is actually higher than the measured DC transconductance, which indicates that the present HEMTs are not as dispersive as earlier ones.

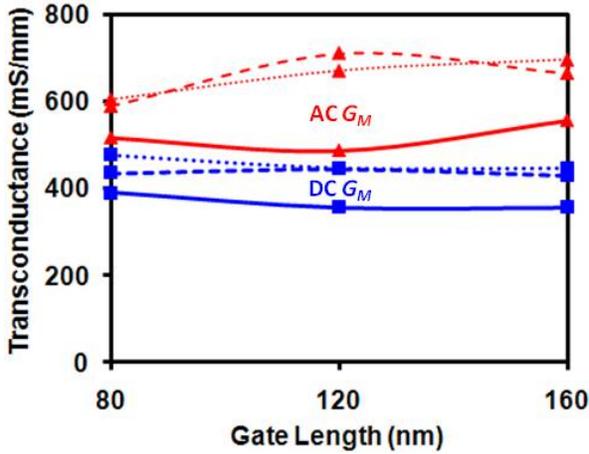

Fig. 4. Measured (■) DC $G_M$ vs. extracted (▲) AC $G_M$ of present AlN/GaN HEMTs. $W_G$ = (· · ·) 50 × 2 μm, (- - -) 100 × 2 μm, or (——) 150 × 2 μm.

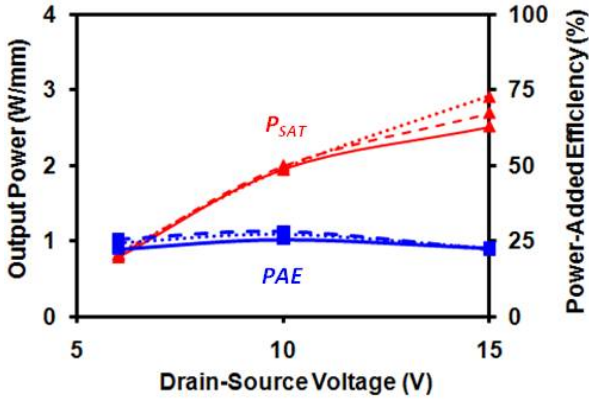

Fig. 5. Measured (▲) $P_{SAT}$ and (■) $PAE$ of present AlN/GaN HEMTs. $L_G$ = (· · ·) 80 nm, (- - -) 120 nm, or (——) 160 nm. $W_G$ = 150 × 2 μm. $V_{GS}$ = −1 V. $V_{DS}$ = 7 V. $f$ = 12 GHz.

## IV. LARGE-SIGNAL CHARACTERISTICS

As listed in Table I, although both saturated output power density $P_{SAT}$ and power-added efficiency $PAE$ of the present HEMTs decrease with increasing gate length and width, the dependence is not very strong. This suggests that there is ample room to scale the HEMTs for higher power. Although both $P_{SAT}$ and $PAE$ were measured at a much higher frequency than that of [13], their values were comparable. In both cases, $P_{SAT}$ ≈ 2.5 W/mm is quite reasonable for $V_{DS}$ ≈ 15 V, $V_{KNEE}$ ≈ 2 V, and $I_{MAX}$ ≈ 1 A/mm, where $I_{MAX}$ is the maximum open-channel current. Fig. 5 shows that $P_{SAT}$ increases with $V_{DS}$ sub-linearly and increasing $V_{DS}$ beyond 15 V would degrade the HEMTs by generating and trapping hot electrons under the gate [29]. By contrast, $PAE$ is independent of $V_{DS}$ and is in general low for these relatively small devices. Fig. 6 shows that low $PAE$ is probably caused by soft gain compression due to buffer leakage.

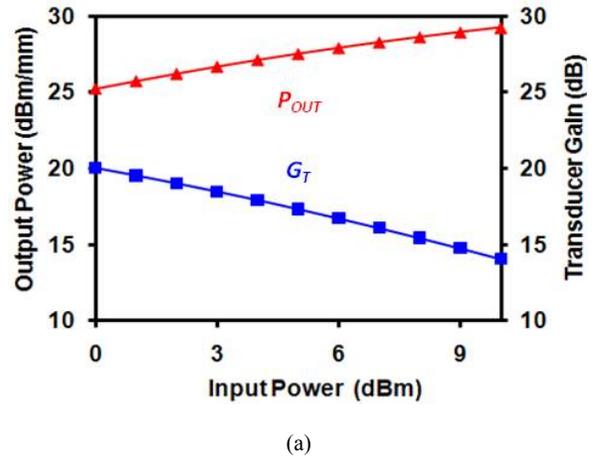

(a)

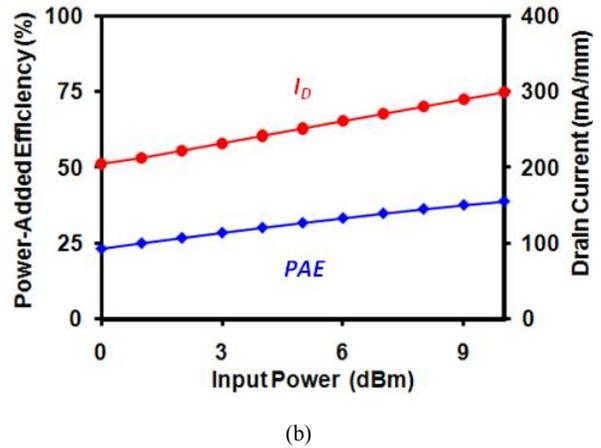

(b)

Fig. 6. Measured (a) (▲) $P_{OUT}$ and (■) transducer gain $G_T$, and (b) (●) drain current $I_D$, and (♦) $PAE$ and for present AlN/GaN HEMT. $L_G$ = 160 nm. $W_G$ = 150 × 2 μm. $V_{GS}$ = −1 V. $V_{DS}$ = 7 V. $f$ = 12 GHz.

## V. CONCLUSION

For the gate lengths and widths investigated in this work, the AlN/GaN HEMTs appear to scale well and consistent with the literature. This suggests that there is room for further scaling to achieve even higher frequency and higher power performance. Although the present output power density appears to be reasonable, it can be improved by evening out the field distribution between the gate and the drain to minimize the generation of hot electrons, and by improving the quality of the gate stack to avoid trapping any hot electron generated. Similarly, although short-channel effects appear to be under control, they can be further suppressed by reducing buffer leakage.

ACKNOWLEDGEMENT

This work was supported in part by the US National Aeronautics and Space Administration under contract no. NNX09C76C.